\documentstyle[12pt,aaspp4]{article}
\begin{document}

\newcommand{\up}[1]{\ifmmode^{\rm #1}\else$^{\rm #1}$\fi}
\newcommand{\zdot}{\makebox[0pt][l]{.}}
\newcommand{\upd}{\up{d}}
\newcommand{\uph}{\up{h}}
\newcommand{\upm}{\up{m}}
\newcommand{\ups}{\up{s}}
\newcommand{\arcd}{\ifmmode^{\circ}\else$^{\circ}$\fi}
\newcommand{\arcm}{\ifmmode{'}\else$'$\fi}
\newcommand{\arcs}{\ifmmode{''}\else$''$\fi}

\title{The Araucaria Project: Dependence of mean K, 
J, and I absolute magnitudes of red clump stars on 
metallicity and age
\footnote{Based on  observations obtained with the VLT
and NTT telescopes at the European Southern Observatory as 
a part of proposal 68.D-0352}}

\author{G. Pietrzy{\'n}ski}
\affil{Universidad de Concepci{\'o}n, Departamento de Fisica, Casilla 160--C, 
Concepci{\'o}n, Chile}
\affil{Warsaw University Observatory, Al. Ujazdowskie 4,00-478, Warsaw, Poland}
\authoremail{pietrzyn@hubble.cfm.udec.cl}
\author{W. Gieren}
\affil{Universidad de Concepci{\'o}n, Departamento de Fisica, Casilla 160--C, 
Concepci{\'o}n, Chile}
\authoremail{wgieren@coma.cfm.udec.cl}
\author{A. Udalski}
\affil{Warsaw University Observatory, Al. Ujazdowskie 4,00-478, Warsaw, Poland}
\authoremail{udalski@astrouw.edu.pl}
\begin{abstract}
As a part of our ongoing Araucaria Project on the improvement of stellar distance indicators, we
 present deep near-infrared JK imaging of several fields in four Local Group galaxies:
LMC, SMC, and the Carina and Fornax dwarf galaxies. These data were obtained under excellent seeing
conditions at the ESO VLT and NTT telescopes. We determine the mean red clump star magnitudes in
the J and K bands in the four galaxies. A comparison of the extinction-corrected K-band red clump star magnitudes
with the tip of the red giant branch magnitude, the mean RR Lyrae star V-band magnitude, and the mean
K-band magnitude of Cepheid variables at a period of 10 days (for the LMC and SMC) strongly suggest
that the red clump star absolute K-band magnitude has a very low (if any) dependence on metallicity
over the broad range of metallicities covered by our target galaxies. This finding is in contrast to
the mean I-band and J-band red clump star magnitudes, which do have a clear metallicity dependence
which we calibrate from our data. Excellent agreement with the former calibration of the red clump
I-band magnitude dependence on metallicity of Udalski is found from our new data. We use the Galactic
cluster K-band red clump star data of Grocholski and Sarajedini to demonstrate that the K-band
red clump star absolute magnitude also has very little (if any) dependence on age over an age
range of about 2-8 Gyr. The present study therefore provides clear evidence that the mean K-band magnitude
of red clump stars is an excellent distance indicator, with very small (if any) population corrections
to be applied over a large range in metallicity and age. Our findings imply that present-day population
corrections calculated from models are only accurate at a +/- 0.15 mag level, which is a great achievement
in itself but not accurate enough for high-precision distance scale work.

We determine the distances to all our target galaxies from the K-band red clump magnitude, with very
small statistical uncertainties. Comparing these distances to those coming from the observed mean I-band magnitudes
of the red clump stars, we find evidence that there is likely to be a problem in the
photometric calibration of the local, solar neighborhood red clump star K-band or I-band magnitudes
which amounts to some 0.2 mag. A re-determination of the absolute photometric calibration of the
Hipparcos-observed nearby red clump stars seems necessary to resolve this problem and put the
derivation of absolute distances to Local Group galaxies from their red clump stars on a firmer basis.

\end{abstract}

\section{Introduction}
We have recently started on a long-term observational programme  
called the Araucaria project, which intends to check on the systematics 
and true capabilities of several stellar methods for distance measurement, using our 
basic strategy to apply  
such stellar techniques of distance determination on a common sample of galaxies with widely
 different environmental properties, 
and to compare the results. 
This should eventually lead to a clear picture about the true capabilities of the different 
methods, and their dependence on environmental properties like metallicity
and age, and will help us to select, calibrate and apply those techniques which give
the most accurate results. A more detailed description of the Araucaria project can be found 
in Gieren et al. (2001). 

The helium-burning intermediate-age red clump stars are 
particulary interesting and important in our studies. The main advantage in
using these stars as standard candles is their small range of observed 
brightnesses, in any given photometric band, and the fact that they are very numerous in 
the solar neighborhood. About of one thousand red clump stars have trigonometric 
parallaxes measured by the Hipparcos satellite with an accuracy of better than 10 \%, 
which made it possible to calibrate this method with an unprecedented accuracy
(Paczy{\'n}ski and Stanek 1998, Alves 2000). 

The main concern about using any object as a distance indicator is related to 
possible population effects which might affect its brightness in different 
environments. Observations of red clump stars in clusters in the Magellanic Clouds indicate
that the dependence of the mean I-band magnitude on age is practically negligible
(within $\pm$ 0.05 mag) for the age range 2-9 Gyr (Udalski 1998). Detailed studies 
of the dependence of $<I>^{\rm RC}$ on metallicity were performed using a large
set of nearby red clump stars with spectroscopic metallicity determinations
obtained by Mc Williams (1990). Within the range of metallicities covered by the nearby red clump stars 
(-0.5 $<$ [Fe/H] $<$ 0.1) the slope of the brightness-metallicity relation in the I band
turned out to be a small, but non-negligible 0.14 mag/dex (Udalski 2000a).

Important progress was made by Alves (2000) who calibrated the absolute
magnitudes of a large sample of Hipparcos-observed nearby red clump stars. 
An outstanding and obvious advantage of using the K band is that it practically 
eliminates any dependence of the obtained results on the adopted reddening.
Alves (2000) also showed that the K-band magnitude of red clump stars 
does not seem to depend, in any significant way, on metallicity in the range from -0.5 to 0.0 dex. 
The results of the very recent work of Grocholski and Sarajedini (2002) 
also suggest that the mean magnitude of red clump stars in the K band 
may be  an excellent standard candle, and therefore the step 
of taking the method to the near-infrared wavelengths is of a profound importance.

Encouraged by these first results we decided to perform deep 
near-infrared imaging of selected fields in four Local Group galaxies, the LMC, SMC, and the
Carina and Fornax dwarf spheroidals,
in order to study in more detail how the mean K and J absolute magnitudes of red clump stars 
depend on their metallicities and/or ages. The very broad range of these environmental 
properties covered by our target galaxies was expected to be very useful for this purpose.
In this paper, we are reporting on the results of this study.

\section{Observations and Calibrations}
The data presented in this paper were collected with the ESO/VLT telescope equipped 
with the ISAAC infrared camera. In our chosen setup, the field of view was about 2.5 $\times$ 2.5 arcmin
with a scale of 0.148 arcsec/pixel. The observations were performed 
 on two photometric nights ( 9 and 10 November 2001). 
The seeing was about 0.5 and 1 arcsec during the first and second night, 
respectively. The J and Ks observations for two fields in Fornax, and two fields 
in Carina were secured on the first night. In addition, Ks images were obtained 
for a third field in the Carina galaxy in this first night of excellent seeing. During the second night, J and Ks 
imaging was performed for two new fields in Carina, and two additional 
fields in Fornax. We also obtained J-band data for the third field 
in Carina, observed in Ks on the first night. Table 1 gives 
information about the coordinates of our chosen fields, and the number of 
red clump stars found in each of them.

\begin{deluxetable}{c c c c c}
\tablecaption{Observed fields}
\tablehead{
\colhead{Field} & \colhead{RA (J2000)} & \colhead{DEC (J2000)} 
& \colhead{$N_{RC}$} \\
}
\startdata
Fornax FI  & 02:40:08.7& -34:11:28.0 & 82\\
Fornax FII &02:40:30.8& -34:25:13.5 & 204\\
Fornax FIII&02:39:32.9& -34:34:24.2 & 291\\
Fornax FIV &02:39:35.8& -34:25:37.4 & 242\\
Carina FI  &06:41:26.9& -50:58:33.4 & 46\\
Carina FII &06:41:14.3& -51:00:32.6 & 39\\
Carina FIII&06:40:59.8& -50:59:11.4 & 47\\
Carina FIV &06:41:57.8& -50:54:33.6 & 31\\
Carina FV  &06:42:09.3& -50:55:09.1 & 28\\
LMC FI & 05:33:39.5 & -70:09:36.5   & 464 \\
LMC FII& 05:33:53.8 & -70:04:01.9   & 494 \\
LMC FIII&05:22:42.9 & -69:31:46.2   & 542 \\
SMC FI  &00:43:44.5  & -73:17:56.1   & 252 \\
SMC FII  &00:44:05.2 & -73:20:49.2  & 219 \\ 
\enddata
\end{deluxetable}

In order to account for the frequent sky level variations in the infrared domain, 
especially in the K band, the observations were performed with a jittering 
technique. In the Ks filter, we did 5 consecutive 12 s integrations 
(DITs) in any given 
sky position before moving the telescope by about 20 arcsec to a different 
position. 24 and 78 such jittering positions obtained for the Carina and Fornax 
galaxies resulted in a total net exposure time of 24,  and 78 minutes, respectively.
The observations in the J band were made in a similar fashion. The total net
exposure times for Fornax and Carina in this filter were 480 s and 240 s 
during the first night, and 800 and 360 s during the second night.
 
In order to accurately transform our data to the standard system, we secured 
8 and 6  observations of standard stars from the UKIRT system ( Hawarden et al. 2001),
on the first and second night, respectively, at a variety of airmasses. 
These standard stars were chosen to span a broad range in colors bracketing the colors 
of the red clump stars in our fields. 
Aperture photometry of the standard stars was performed with the DAOPHOT program using 
apertures with radii of 16 pixels. 
Transformations coefficients were derived for each night. The accuracy 
of the zero points was calculated to be  about 0.025 mag. 

The data reductions were performed in an exactly analogous manner as in Pietrzy{\'n}ski and Gieren (2002).
In brief, the sky was subtracted from the images using a two-step process implying the masking of stars 
with the xdimsum IRAF  package. After flatfielding and stacking into final images 
the PSF photometry was performed with the DAOPHOT and ALLSTAR programs, following 
the procedure described in detail in Pietrzy{\'n}ski, Gieren and Udalski (2002). 
Aperture corrections were derived using about 10-20 relatively bright and isolated 
stars, after iteratively removing all stars from their neighborhood. The final 
aperture corrections were adopted as the median taken from all measurements and 
typically had a rms scatter of about 0.015 mag.

A more detailed description of the observations, data reduction and calibration 
procedures together with a discussion of the morphology of near-infrared color-magnitude diagrams
will be presented in Pietrzy{\'n}ski, Gieren and Udalski (2003, in preparation). 

As part of the same project, we observed 10 different fields in the 
LMC and SMC on 3 nights with the ESO NTT telescope in the J and Ks filters.
We have already used a part of these data to derive the distance to the LMC using 
K-band photometry of red clump stars (Pietrzy{\'n}ski and Gieren, 2002). The detailed 
description of the complete set of data collected for the Magellanic Clouds will be presented 
in a separate paper (Pietrzy{\'n}ski, Gieren and Udalski, 2003 in preparation). 
Here we  use photometry obtained with the ESO NTT on 2 nights for 3 fields in the LMC and 2 fields in the SMC, 
to study the magnitudes of red clump stars in these galaxies.
 Basic information on the  data is given in Table 1.
Observations, reductions and calibrations of these Magellanic Cloud data were
performed in the same manner as for the data obtained for the Fornax and Carina dwarf
galaxies. The accuracy of the derived zero points for the 2 NTT observing nights  was better than 
0.03 mag.
It is worthwhile noticing that the instrumental systems at both the NTT and VLT telescopes 
are identical, which obviously helped us to obtain very homogenous data, for all four galaxies 
discussed in this paper.

\section{Mean K- and J-band red clump magnitudes}

The near-infrared K vs J-K color-magnitude diagrams around the red clump region are shown
in Fig. 1 for our four target galaxies.
The relatively large number of red clump stars found in our 4 fields 
in Fornax (see Table 1) allowed us to derive their mean K-band 
 magnitudes individually for each field in this galaxy.
We did this by selecting the stars with K-band magnitudes in the range 
from  18.3 to 20.3 mag, and J-K colors in the range from 0.3 to 0.85 mag
and deriving the histogram of their K-band magnituds in bins of 
0.06 mag for each field. Then function (1), which 
consists of a Gaussian component representing the red clump stars 
distribution and a second order polynomial function approximating the 
stellar background was fitted to the data, following a procedure originally
introduced by Paczy{\'n}ski and Stanek (1998): 

$$n(K)=a+b(K-K^{\rm max})+c(K-K^{\rm max})^2+
\frac{N_{RC}}{\sigma_{\rm RC}\sqrt{2\pi}} \exp\left[-\frac{(K-K^{\rm
max})^2}{2\sigma^2_{\rm RC}}\right]\eqno(1)$$ \\

The same procedure was applied for deriving the mean red clump star
magnitudes in the J band. The results are presented
in Table 2. It can be appreciated that the four independent measurements of the mean 
J- and K-band red clump magnitudes obtained on the two different observing nights 
are in very good agreement. We therefore 
decided to merge the data in order to improve the statistics.
The fits of function (1) to the histograms of the K and J magnitudes 
of red clump stars from the combined data are displayed in Fig. 2. The corresponding results 
of $<K>$ = 19.215 $\pm$ 0.013 and $<J>$ = 19.687 $\pm$ 0.014 were finally 
adopted as the mean K- and J-band apparent magnitudes of red clump stars in the Fornax dwarf galaxy. 

Using the  same approach we derived $<J>$ and $<K>$  for red 
clump stars in the FI and FII fields in the SMC, and the FIII field in the LMC. 
Appropriate data for the fields FI and FII in the LMC were adopted from Pietrzy{\'n}ski
and Gieren (2002). Again, very good agreement of the red clump star magnitudes for the different
fields in LMC and SMC, respectively, was found, justifying the procedure to merge the
data of the individual fields. 
This led to the following adopted red clump star magnitudes:
$<K>_{\rm LMC}$ = 16.894 $\pm$ 0.007, $<J>_{\rm LMC}$ = 17.472 $\pm$ 0.007, 
$<K>_{\rm SMC}$ = 17.346 $\pm$ 0.018, $<J>_{\rm SMC}$ = 17.857 $\pm$ 0.020 mag.  
The corresponding fits are displayed in Fig. 3.

It is worthwhile noticing that our $<K>_{\rm LMC}$ on the UKIRT system 
is in very good agreement with $<K>_{\rm LMC}$ = 16.974 derived by Alves et al. (2002),
on the Koorneef system (see Pietrzynski and Gieren 2002, for more details). 

Our fields in the Magellanic Clouds are located so close (within about 
0.5 degree) to the centers of these galaxies that any geometrical 
corrections to the red clump star magnitudes of the different fields are expected to be very small. 
We checked this by applying the geometrical 
 model of van der Marel et al. (2002) for the LMC, and that of
Groenewegen (2000) for the SMC. The results were consistent with 
our previous estimations of the red clump star mean magnitudes 
to within 0.01 mag (e.g. clearly less than the accuracy of our zero 
point measurements), justifying our procedure to merge the data of our individual
fields in the LMC and SMC without previously applying corrections for the geometrical
structures of the Clouds.

In the case of Carina, the number 
of red clump stars we found is not large enough to derive mean red clump stars magnitudes 
for the individual fields. However, it should be noted that all our five selected
 fields are located very close each other (the distance between them 
is smaller than 10 arcmin), so we decided to merge the observations for the 2 nights and different fields 
and to derive the mean K and J magnitudes of the red clump stars 
in the same manner as we did for Fornax and the Magellanic Clouds. 
The  fit of function (1) to the combined data for Carina yielded
as our adopted, final red clump star mean magnitudes for this galaxy 
$<K>$ = 18.533 $\pm$ 0.015 and $<J>$ = 18.970 $\pm$ 0.014 mag. 
The fits in the case of Carina are shown in Fig. 2.

\begin{deluxetable}{c c c c c c}
\tablecaption{Mean J and K magnitudes of red clump stars in Fornax, Carina, LMC and SMC}
\tablehead{
\colhead{Field} & \colhead{$<K>$} & \colhead{$\sigma_{\rm K}$}
& \colhead{$<J>$} & \colhead{$\sigma_{\rm J}$}\\
}
\startdata
Fornax FI  & 19.26 & 0.05  & 19.72 & 0.05\\
Fornax FII & 19.227 & 0.016 & 19.691 & 0.017 \\
Fornax FIII& 19.213 &  0.016 & 19.680 & 0.016\\
Fornax FIV & 19.192 & 0.021 & 19.668 & 0.019\\
Fornax all data & 19.215 & 0.013 & 19.687 & 0.014\\
Carina night 1  & 18.540 & 0.015 & 18.969  & 0.023\\
Carina night 2  & 18.51 & 0.04 & 18.99 & 0.05\\
Carina all data & 18.533 & 0.015 & 18.970 & 0.014\\
LMC FI     & 16.893 & 0.013 & 17.508 & 0.009 \\
LMC FII    & 16.898 & 0.010 & 17.523 & 0.016 \\
LMC FIII   & 16.892 & 0.010 & 17.450 & 0.008 \\
LMC all data & 16.894 & 0.007 & 17.472 &  0.007 \\
SMC FI     & 17.369 & 0.026 & 17.866 & 0.020 \\
SMC FII    & 17.339 & 0.020 & 17.84  & 0.03 \\
SMC all data & 17.346 & 0.018 & 17.857 & 0.020\\
\enddata
\end{deluxetable}

\section{Comparison with other distance indicators}
In Table 3, we give the reddening-corrected $<K_{0}>$ magnitudes of the red clump stars in the LMC, 
SMC, Carina and Fornax. For comparison, we also give  
the I-band magnitudes of the TRGB and the mean V-band magnitude of RR Lyrae stars in these galaxies, 
taken from the literature. The magnitudes for the TRGB and the RR Lyrae 
stars in the Magellanic Clouds and Carina dwarf galaxy were adopted from Udalski 
(2000b). $<I^{\rm TRGB}>$ for Fornax was taken from 
Bersier (2000).  The reddenings towards the Fornax 
and Carina galaxies were derived from the Schlegel, Finkbeiner and Davis (1998)
extinction maps, yielding $E(B-V)_{\rm Car}$ = 0.06 $\pm$ 0.01 mag, and
$E(B-V)_{\rm For}$ = 0.03 $\pm$ 0.01 mag, respectively. 
Extinction corrections to our fields in the 
Magellanic Clouds were applied by adopting the appropiate values of E(B-V) from the
OGLE II extinction maps in the LMC (Udalski et al. 1999a ; $E(B-V)_{\rm FI, FII}$ = 0.152, 
 $E(B-V)_{\rm FIII}$ =0.115 mag)
and in the SMC (Udalski et al. 1999b ; $E(B-V)_{\rm FI, FII} = 0.089$) using the 
reddening law as given by Schlegel et al. (1998).


Finally, the mean $<K_{0}>$ magnitudes of Cepheids at a pulsation period
of 10 days in the Magellanic 
Clouds were calculated using the relations provided by Groenewegen (2000). These are also
given in Table 3.

For our further discussion, we adopt -0.5 dex for the metallicity of intermediate-age populations in the LMC 
(Bica et al. 1998, Smecker-Hane et al. 1999), and -1.0 dex for the SMC 
(see Udalski 2000b for a discussion).
For Fornax, we assume [Fe/H] = -1.0 (Saviane, Held and Bertelli, 
2000; Tolstoy et al. 2002). From spectroscopic metallicity determinations of 52 
stars in Carina,  Smecker-Hane et al. (1999) derived a mean metallicity of 
red giants in this galaxy of -2.0 dex. More recently Tolstoy et al. (2002), 
using high resolution spectra obtained with UVES at the VLT for 5 stars, 
 argued that 
the mean metallicity of red giant stars in the Carina dwarf galaxy is less metal-poor and about -1.6 dex.
We adopt the mean from these two determinations for Carina, e.g. -1.8 dex.

\begin{deluxetable}{c c c c c}
\tablecaption{Mean observed magnitudes of major distance indicators in our target galaxies}
\tablehead{
\colhead{Galaxy} & 
 \colhead{$<K_{0}^{\rm RC}>$} & \colhead{$<I_{0}^{\rm TRGB}>$}
& \colhead{ $<K_{0}^{\rm Cep,10d}>$ } & \colhead{ $<V_{0}^{\rm RR Lyr}>$ } \\
& mag & mag  & mag  & mag\\
}
\startdata
LMC    & 16.844$\pm$0.007  & 14.33$\pm$0.02 & 12.79$\pm$0.02  & 18.91$\pm$0.01 \\
SMC    & 17.313$\pm$0.018  & 14.83$\pm$0.02 & 13.28$\pm$0.02  & 19.44$\pm$0.02 \\
Carina & 18.511$\pm$0.015  & 16.03$\pm$0.05 &      --         & 20.61$\pm$0.03 \\
Fornax & 19.204$\pm$0.013  & 16.65$\pm$0.05 &      --         &     --      \\
\enddata
\tablecomments{$<V_{0}^{\rm RR Lyr}>$ at ${\rm [Fe/H]}_{\rm LMC}$ = -1.6 dex (Alcock et al. 1996)}
\end{deluxetable}
 
It was shown by empirical investigations
(Lee, Freedman and Madore 1993; Udalski 2000b) that the I-band magnitude of the
TRGB is independent of population effects in the range of   
ages and metallicities covered by our 4 target galaxies. 
Moreover, comparisons between TRGB
distances and Cepheid distances in many galaxies (Lee, Freedman and Madore 1993;
Kennicutt et al. 1998; Ferrarese et al. 2000; Udalski 2000b) showed excellent agreement 
between these two distance indicators. Therefore, by 
studying the difference of $<K_{0}^{\rm RC}>$ and $<I_{0}^{\rm TRGB}>$ as a function of metallicity
and age
we should be able to detect any population effects on $<K_{0}^{\rm RC}>$.
Fig. 4 shows such a comparison as a function of metallicity. It is appreciated from this diagram that
there is clearly no systematical trend in the difference of the red clump K-band magnitude and the TRGB
I-band magnitude with metallicity. A linear regression leads to 
$<K_{0}^{\rm RC}>$ - $<I_{0}^{\rm TRGB}>$ = 2.51, with a dispersion of only
0.04 mag.
An analogous comparison of the red clump star mean K-band magnitude with $<V_{0}^{\rm RR Lyrae}>$,
where the RR Lyrae mean V-band magnitude has been corrected for a metallicity slope of 0.18  (see
Udalski 2000b, and references therein) yields a very similar result
(see Fig. 4, lower panel). 

Another, independent test of a possible trend of $<K_{0}^{\rm RC}>$ with metallicity
can be done using the 2MASS near-infrared photometry of 
Cepheids in the Magellanic Clouds (Groenewegen 2000). Here it is worthwhile noticing 
that there is practically no difference between the 2MASS Ks-band and 
the K-band magnitudes in the UKIRT system to which our data have been transformed
(Carpenter 2001). Using the P-L relations published by Groenewegen (2000), 
we derive $<K_{0}^{\rm RC}>$ - $<K_{0}^{\rm Cep, P=10 d}>$ = 4.05 $\pm$ 0.02 for the LMC,
and 4.03$\pm$0.03 for the SMC, again indicating that this magnitude difference does not depend
on metallicity (at least in the metallicity range covered by LMC and SMC).
 
All our comparisons seem to indicate that {\it $<K_{0}^{\rm RC}>$ does not, or very little, depend on population 
effects in the very broad metallicity range (-1.8 $<$ [Fe/H] $<$ -0.5) 
and the very different star formation histories covered by our four target galaxies}.  
This is in very good agreement with the results of Alves (2000) who also found 
that the $<K_{0}^{\rm RC}>$ absolute magnitude of red clump stars in the solar neighborhood does not show
any significant trend with metallicity in the range from -0.5 to 0.0 dex.

In principle, the dependence of $<K_{0}^{\rm RC}>$ on metallicity and age can also be checked 
with star clusters containing sufficient red clump stars, and spanning a range in ages and metallicities. 
Unfortunately, the number of 
clusters with available accurate infrared photometry is very low at the present time.
A first attempt to study the behaviour  of $<K_{0}^{\rm RC}>$ in a sample of 
Galactic clusters was recently performed by Grocholski and Sarajedini (2002). 
They took infrared data for 14 open clusters from the 2MASS database and 
found an over-all agreement of the data with model predictions about the dependence of the K-band
red clump star magnitude on age and metallicity. 
However, the accuracy of their results was clearly limited by the low number of red clump stars in the
studied clusters, and the distance uncertainies of the clusters inherent in the MS fitting technique
of distance determination. The  
typical accuracy of $M_{\rm K}^{\rm RC}$ they were able to achieve was about 0.11-0.2 mag. 
Still, at least a crude check on the possible age dependence of the red clump star K-band absolute magnitude
is possible from the Grocholski and Sarajedini cluster data. We did this by selecting
  clusters having ages in the 
range from 2 Gyr to 9 Gyr (covering the age range of the red clump stars in the target galaxies of this paper).
Three clusters (NGC 6819, M 67 and Be 39) were identified. Their metallicities fall in 
the range covered by the Hipparcos red clump star sample, therefore no dependence  of their 
$M_{\rm K}^{\rm RC}$ on metallicity is expected (Alves 2000). In Fig. 5, we show $M_{\rm K}^{\rm RC}$
for these clusters versus the cluster age. Within the error bars on the absolute magnitudes, no
indication of a systematic trend of the red clump star K-band absolute magnitude with age is visible
in this diagram, suggesting that the red clump star K-band absolute magnitude is not only very insensitive
 to metallicity, but
also to age, over a broad range of ages.

\section{Population dependence of red clump star magnitudes in the I and J bands}
We can now trace out the population dependence of the $<I>_{0}^{\rm RC}$
 and $<J>_{0}^{\rm RC}$  magnitudes very accurately,  studying the relative magnitudes of these 
stars in our target galaxies. The I-band data for the red clump stars in the different galaxies 
were taken from the literature (Udalski et al. 1998, Udalski et al. 2000, Udalski 2000b for
 LMC, SMC and Carina; Bersier 2000 for Fornax) and were dereddened in exactly the
same way as  our infrared data presented in this paper.

As can be seen in Fig. 6, the differences of  $<I>_{0}^{\rm RC} - 
<K_{0}^{\rm RC}>$
and $<J>_{0}^{\rm RC} -  <K_{0}^{\rm RC}>$ are both strongly corelated with metallicity.
Least-squares fits to a straight line give the following slopes: 0.09$\pm$0.02, and 0.13
+-0.02 mag/dex, for the metallicity dependence of $<J>_{0}^{\rm RC}$ and $<I>_{0}^{\rm RC}$, 
respectively. This result is in amazing agreement with the former result of Udalski
(2000a) on the metallicity dependence of the red clump star I-band magnitude, who had
 obtained $<I>_{0}^{\rm RC} = (0.14 \pm 0.04)\times[Fe/H])$ + const, 
from a study of the Hipparcos red clump stars with high-quality spectroscopic metallicity 
determinations in the range from 0 to -0.5 dex.  
The  dependence of  $<J>_{0}^{\rm RC}$ on metallicity is found to be slightly weaker than 
that of $<I>_{0}^{\rm RC}$, as one could expect from the
effective wavelengths of the J-band (1.2 $\mu$) and I-band (0.8 $\mu$).

\begin{deluxetable}{c c c c}
\tablecaption{Mean IJK magnitudes of red clump stars in our target galaxies}
\tablehead{
\colhead{Galaxy} &
 \colhead{$<K_{0}>$} & \colhead{$<J_{0}>$}
& \colhead{ $<I_{0}>$}\\
& mag & mag  & mag  \\
}
\startdata
LMC    & 16.844  & 17.358 & 17.97 \\
SMC    & 17.313  & 17.776 & 18.35 \\
Carina & 18.511  & 18.916 & 19.46 \\
Fornax & 19.204  & 19.659 & 20.24 \\
\enddata
\end{deluxetable}

\section{Discussion}
In Fig. 7, we present the difference between the relative $<I_{0}>$ magnitudes of red clump 
stars derived with respect to the LMC, and corrected for their metallicity dependence, 
and their  $<K_{0}>$ magnitudes (also relative to the LMC). It is clearly seen that in all 
environments {\it except the solar neigborhood} there is excellent agreement between the relative
brightnesses of the red clump stars in these two bands. This finding suggests that there may be
a very significant zero-point error, in the order of 0.2 mag, in either the K-band, or in the I-band
photometry of the local Hipparcos red clump stars (of course it is also possible that
the current local calibrations have significant errors in both bands).

Transforming our K-band data to the Koorneef system with the transformations 
provided by Carpenter (2001), and using the K-band absolute calibration of Alves (2000)
 we derive the 
following true distance moduli from K-band red clump star magnitudes
 to the target galaxies of the present study:
 
18.498$\pm$0.007 mag (LMC)
  
18.967$\pm$0.018 mag (SMC)

20.165$\pm$0.015 mag (Carina dwarf galaxy)
 
20.858$\pm$0.013 mag (Fornax dwarf galaxy)

The quoted uncertainties are the statistical errors of our distance determinations.
The LMC distance result is in excellent agreement with our own former result (Pietrzynski and
Gieren 2002) which was based on smaller number of fields, and with the recent independent distance determinations
to the LMC from red clump star K-band photometry of Alves et al. (2002), and Sarajedini et al.  (2002).

The above distance moduli are to be compared 
with those obtained from I-band data 
corrected for their metallicity dependence: 18.26$\pm$0.07 mag for the LMC (Udalski 2000b); 
18.71$\pm$0.07 for the SMC (Udalski 2000b); 19.96 $\pm$ 0.08 for Carina (Udalski 2000b), and 20.60 for Fornax
(Bersier, 2000).  
 
We then find, as already discussed and shown in Fig. 7, that there is a near-constant difference 
of about 0.2 mag between the distances derived to our target 
galaxies from K-band and I-band magnitudes of red clump stars, using the
local solar neighborhood calibrations of Alves (2000) for the K band and of
Udalski (2000a) for the I band.  It will be of obvious importance to
check on these calibrations in the near future to find out if this way the results
from both bands can be reconciled.

Finally, our current empirical results provide a very strong test on the accuracy 
of the so-called population corrections derived 
from stellar models and assuming a star formation history and chemical 
evolution history for both the solar neighborhood, and a given target galaxy (
Girardi and Salaris, 2002). Using the Girardi and Salaris (2002) corrections for the K band,
 one obtains $d_{\rm SMC}-d_{\rm LMC}  = 0.43$ mag and 
$d_{\rm CAR} - d_{\rm LMC}  = 1.53 $ mag.
If we compare these values with the empirical results for the distance difference between SMC and LMC
 (0.47$\pm$0.03 mag from RC in K without corrections; 0.50$\pm$0.04 mag from TRBG; 
0.52$\pm$0.03 from RR Lyrae stars; 0.49$\pm$0.03 from Cepheids in the K band) 
and for the distance difference between Carina and the LMC (1.70 $\pm$ 0.05 from TRGB; 1.71 $\pm$ 0.03 
from RR Lyrae stars;  and 1.67 $\pm$ 0.03 from the RC in K), 
we can conclude that the current accuracy of the predictions of such models is about 0.1-0.15 mag.
(see also the discussion in Pietrzy{\'n}ski and Gieren, 2002).
Athough it is quite impressive that in spite of the rather poorly known 
SFH and chemical evolution histories in these galaxies one can get results from the models whose
agreement with accurate empirical data is in the order of 0.1-0.15 mag, 
such an approach is clearly of less use for  
accurate distance determinations to nearby galaxies, at the present time.

\acknowledgments
We are grateful to the European Southern Observatory for allocating
observing time to this project with ISAAC on the VLT telescope, and with SOFI
on the NTT telescope. It is
a real pleasure to thank the VLT and NTT teams for their expert support which helped to 
acquire data of the highest possible accuracy. WG gratefully acknowledges 
financial support for this
work from the Chilean Center for Astrophysics FONDAP 15010003. 
He also acknowledges support
from the Centrum fuer Internationale Migration und Entwicklung in
Frankfurt/Germany
who donated the Sun Ultra 60 workstation on which data reduction and
analysis for this project was carried out.   
Support from the polish KBN grant No 2P03D02123 and BST grant for 
Warsaw University Observatory is also acknowledged.

\newpage

\begin{figure}[htb]
\vspace*{18 cm}
\includegraphics{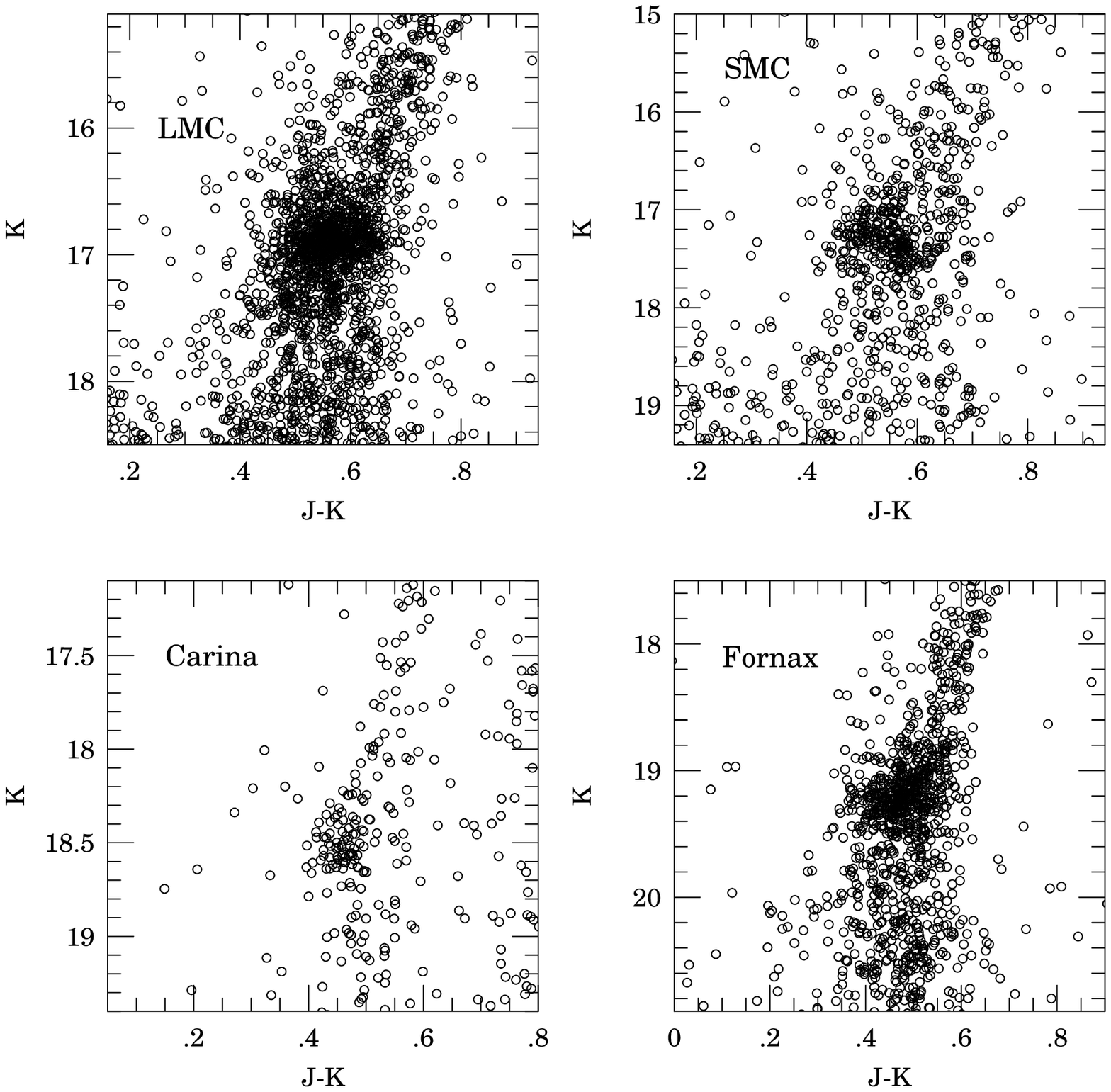}
\caption{K, J-K near-infrared color-magnitude diagrams for the LMC, SMC, Carina and 
Fornax. Data have been combined from the different fields observed in each galaxy (see Table 1).}
\end{figure}

\begin{figure}[htb]
\vspace*{12 cm}
\includegraphics{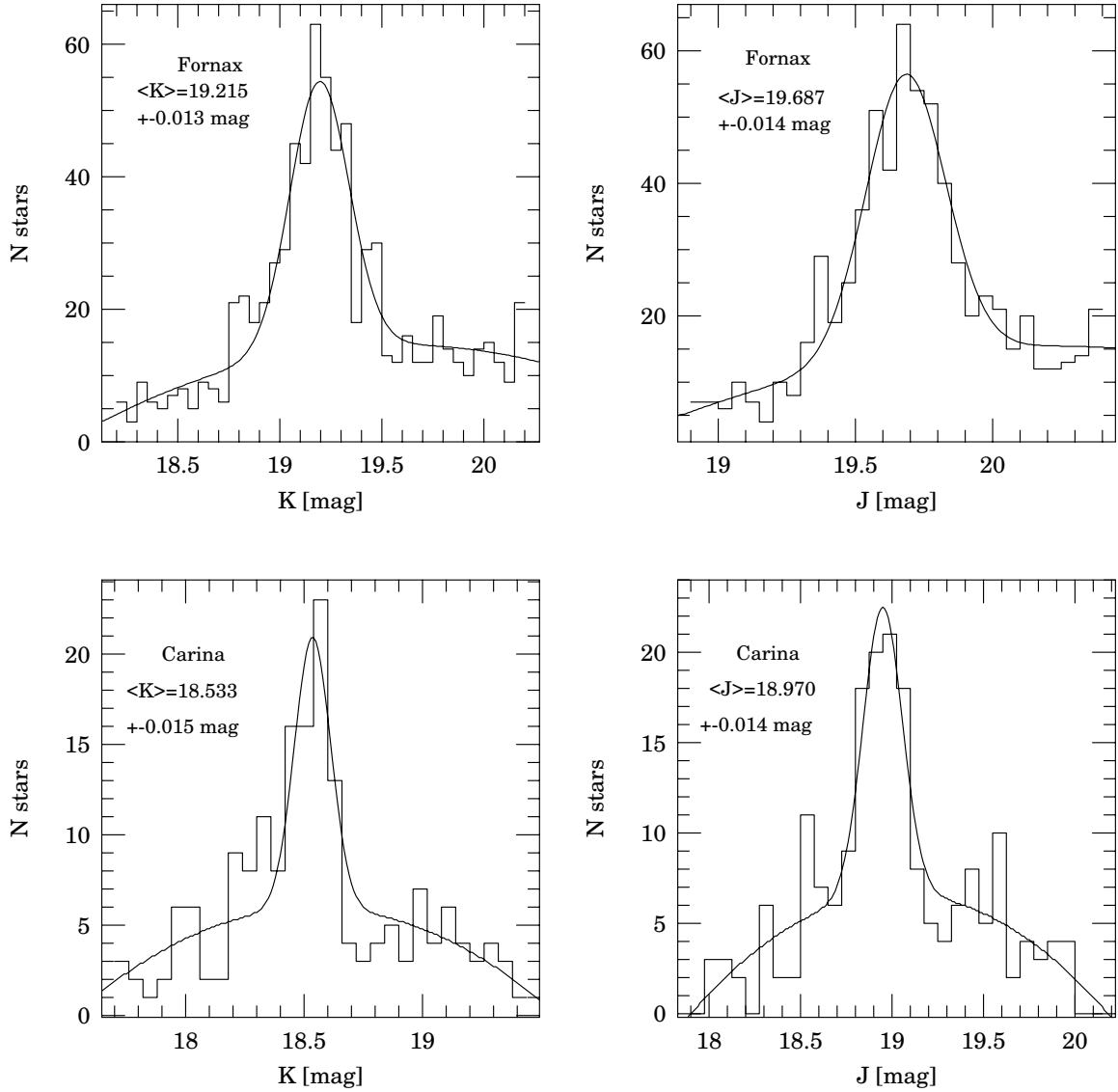}
\caption{The Gaussian and polynomial fit, according to equation (1), applied to 
the stars in a box around the red clump in the Fornax and Carina dwarf galaxies (see text for details). Sharp
and well-defined peaks at for the mean K- and J-band magnitudes are obtained from the data,
with an excellent statistics.
}
\end{figure}

\begin{figure}[htb] 
\vspace*{12 cm} 
\includegraphics{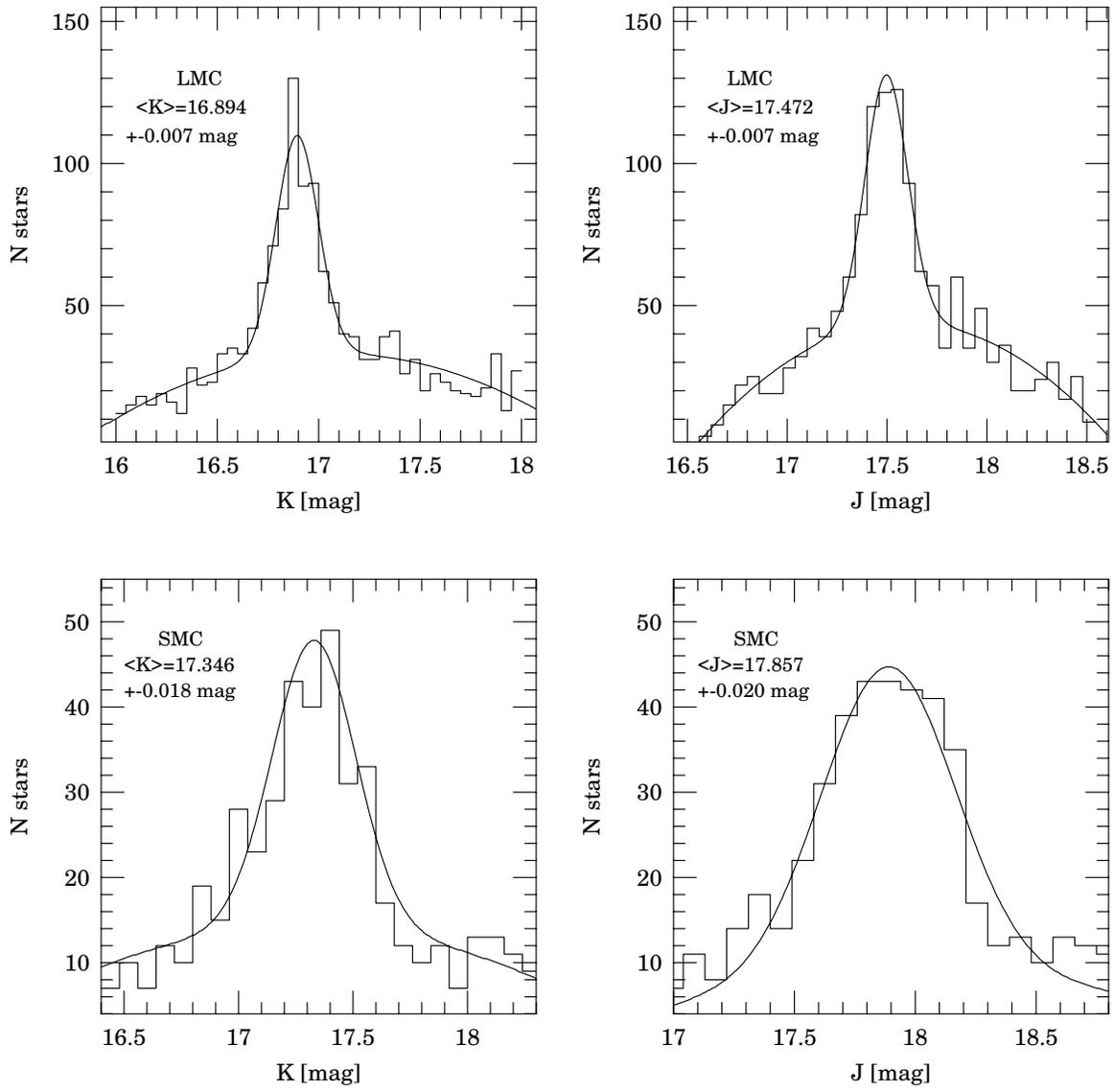}          
\caption{Same as in Fig. 2, but for LMC and SMC.
} 
\end{figure}

\begin{figure}[htb]
\vspace*{10 cm}
\includegraphics{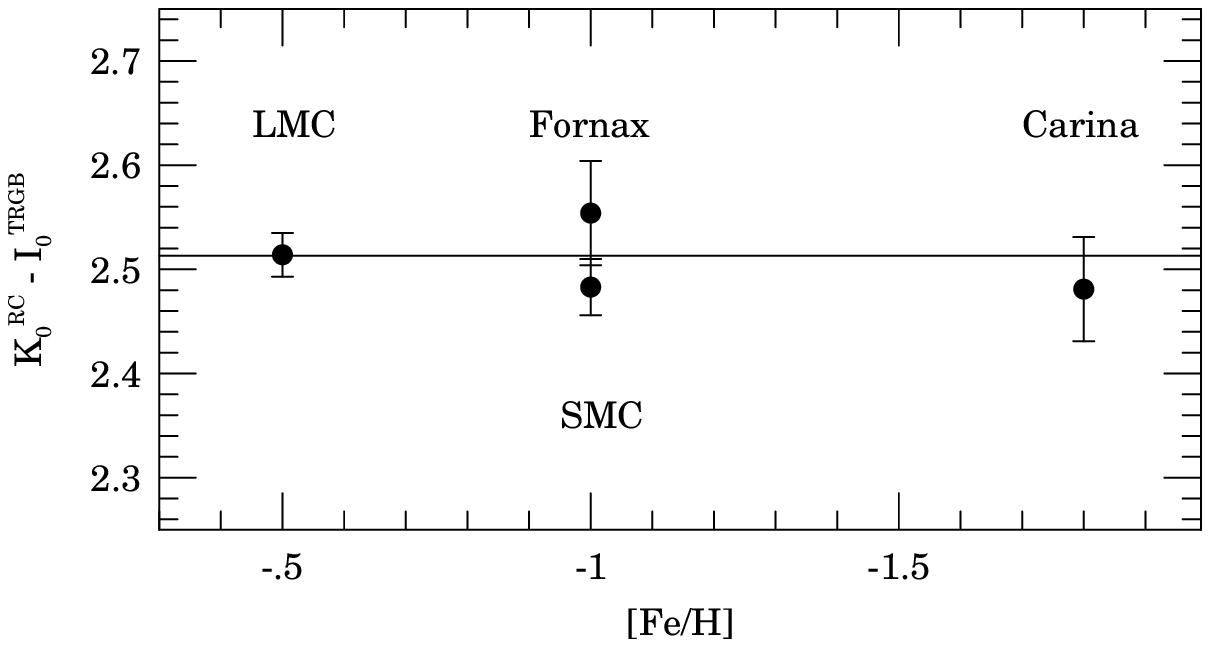}
\includegraphics{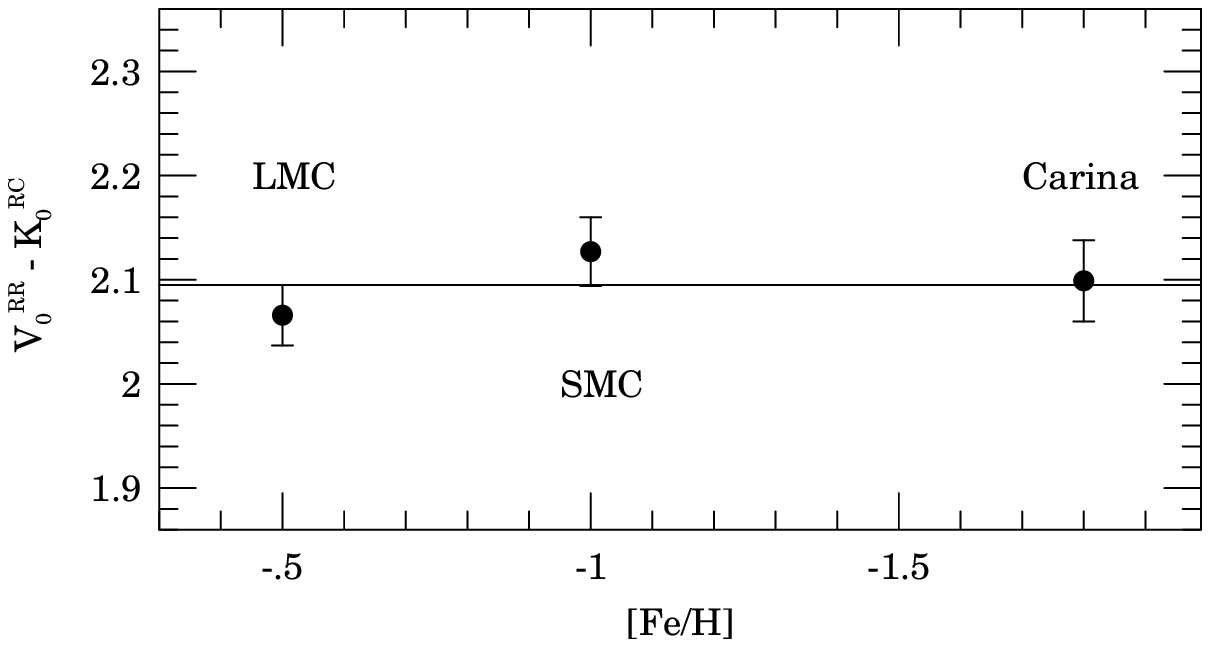}
\caption{Comparison of the red clump star mean K-band magnitude $<K_{0}^{\rm RC}>$ with $<I_{0}^{\rm TRGB}>$
(upper panel), and with $<V_{0}^{\rm RR Lyrae}>$ corrected for a metallicity 
dependence of 0.18*[Fe/H] (lower panel). See text for more details.   
}
\end{figure}

\begin{figure}[htb]
\vspace*{10 cm}
\includegraphics{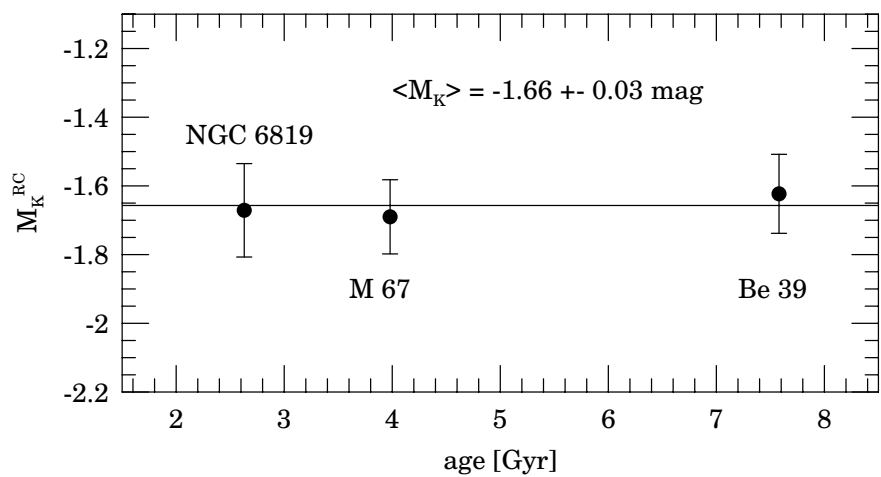}
\caption{ The absolute red clump K-band magnitude ${\rm M}_{\rm K}^{\rm RC}$ from Grocholski 
and Sarajedini (2002) for open clusters having ages in the range from 2 to 9 Gyrs, displayed 
as a function of age. The metallicities of these clusters all lie 
in the range covered by the Hipparcos-observed red clump stars, so no dependence of ${\rm M}_{\rm K}^{\rm RC}$
on metallicity is expected (Alves 2000). The data suggest no significant dependence of 
${\rm M}_{\rm K}^{\rm RC}$ on age over the age range covered by these clusters.
}
\end{figure}

\begin{figure}[htb]
\vspace*{10 cm}
\includegraphics{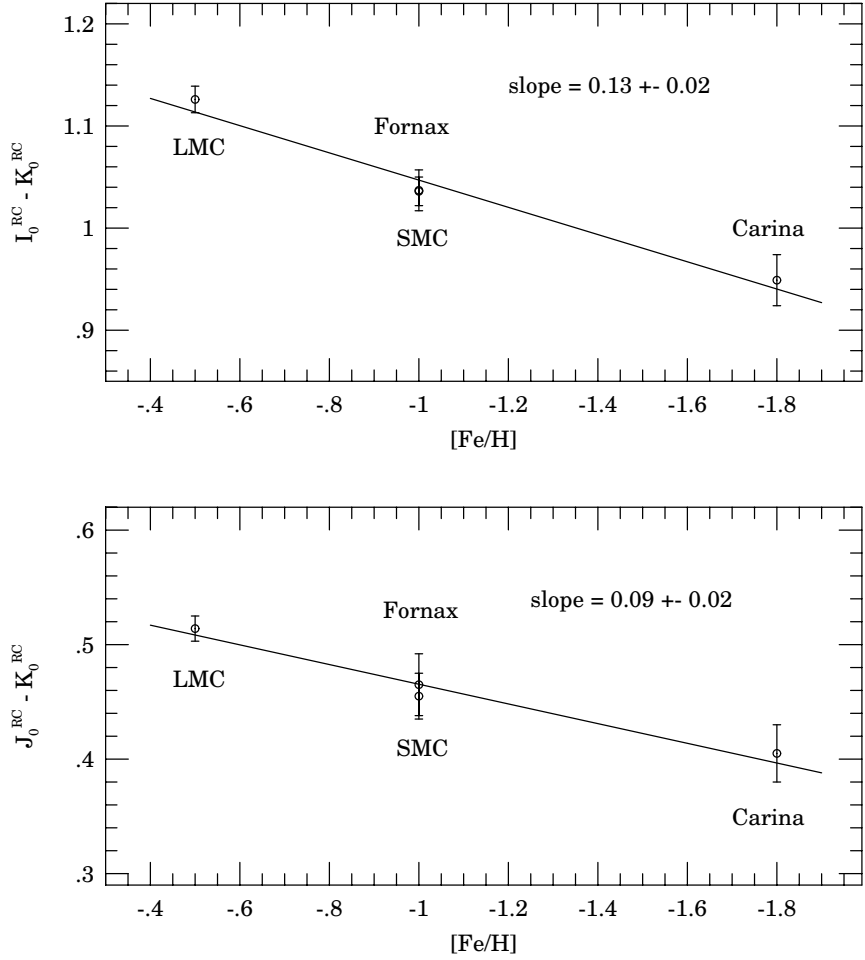}
\caption{ Comparison of $<I_{0}^{\rm RC}>$ and $<J_{0}^{\rm RC}>$ with the metallicity-independent
K-band red clump magnitude.
The observed slopes in these diagrams reflect a clear dependence of both $<I_{0}^{\rm RC}>$ and 
$<J_{0}^{\rm RC}>$ on metallicity. 
}
\end{figure}

\begin{figure}[htb]
\vspace*{10 cm}
\includegraphics{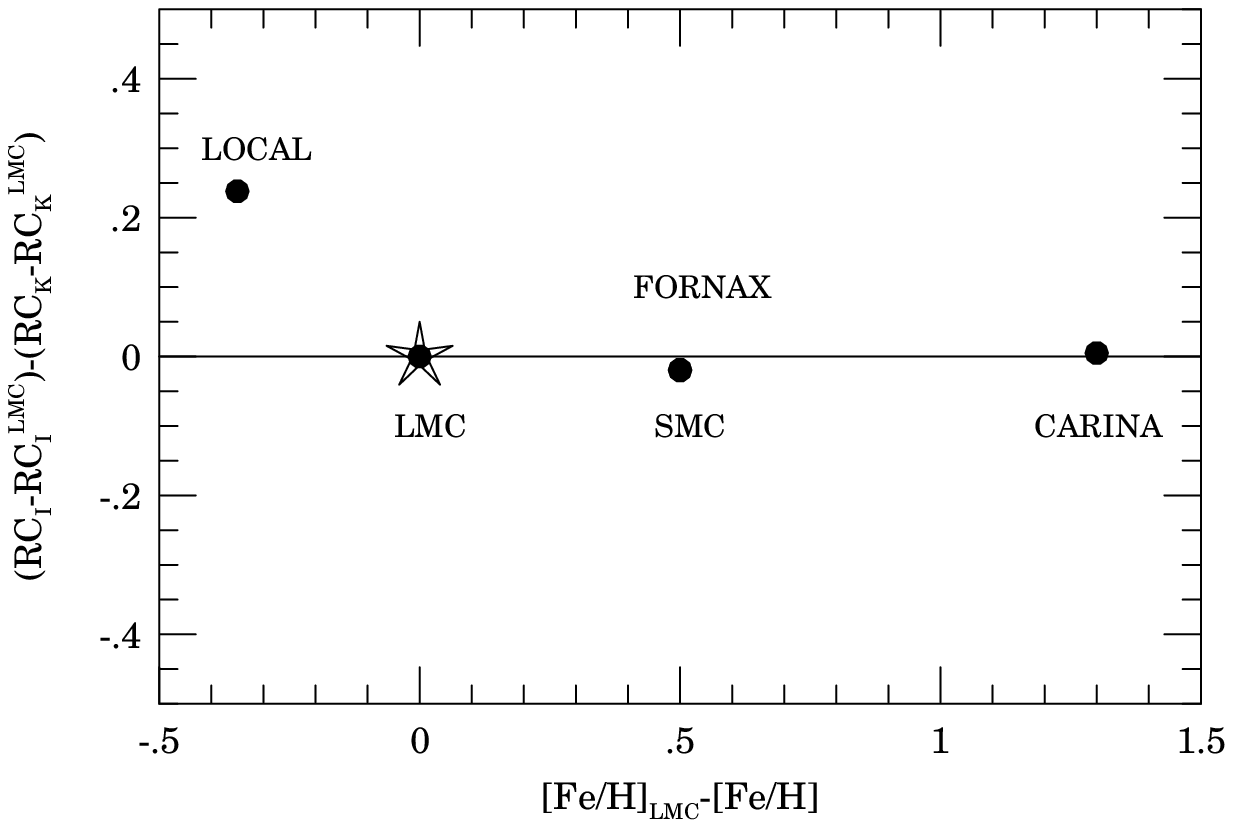}
\caption{ Plot of the difference between the relative (with respect to the LMC) 
$<I_{0}^{\rm RC}>$ (corrected for its metallicity dependence) and 
the corresponding relative $<K_{0}^{\rm RC}>$ against metallicity, for solar neighborhood red clump stars, 
and those in the four target galaxies of the present study. It
can be seen that in all environments this difference is very close to zero,
the exception being the solar neighborhood red clump stars. This suggests that the photometric data for the
Hipparcos-observed red clump stars probably have a zero point error in the order of 
0.2 mag in either the K or the I band.
}
\end{figure}


\begin{references} 
\reference{ircal} Alves, D., et al.  2000, \apj, 539, 732
\reference{irrc} Alves, D., Rejkuba, M., Minniti, D.,  and Cook, K., 2002, \apj, 
573, L51
\reference{BIC} Bica, E., Geisler, D., Dottori, H., Claria, J.J., Piatti, A.E., 
Santos, Jr, J.F.,C.,, 1998, \aj, 116, 723
\reference{david} Bersier, D., 2000, \apj, 543, 23
\reference{transf} Carpenter, J.M., 2001, \aj, 121, 95
\reference{CTRGB2} Ferrarese, L., et al., 2000, \apj, 529, 745
\reference{ara} Gieren, W., Geisler, D., Richtler, T., Pietrzynski, G.,
Dirsch, B., 2001, The Messenger, 106, 15
\reference{dupa} Girardi, L.,  and Salaris, M., 2002, MNRAS, 337, 332
\reference{CEPHII} Grocholski, A.J., and Sarajedini, A., 2002, \apj, 123, 1603
\reference{CEPIR} Groenewegen, M.A.T., 2000, A\&A, 363, 901
\reference{UKIRT} Hawarden, T.G., Leggett, S.K., Letawsky, M.B.,
Ballantyne, D.R., and Casali, M.M., 2001, MNRAS, 325, 563
\reference{CTRGB1} Kennicutt, R.C., Jr., et al., 1998, \apj, 498, 181
\reference{li} Lee, M.G., Freedman, W.L., Madore, B.F, 1993, \apj , 417, 553
\reference{spec} Mc Williams, A., 1990, ApJS, 74, 1075
\reference{rch} Paczy{\'n}ski, B., and Stanek, K.Z., 1998, \apj, 495, L219
\reference{STD} Pietrzy{\'n}ski, G., Gieren, W., 2002, \aj, 124, 2633
\reference{STD} Pietrzy{\'n}ski, G., Gieren, W., and Udalski, A., 2002, \pasp,  
114, 298
\reference{gromdist} Sarajedini, A., Grocholski, A.J., Levine, J., Lada, E., 
2002, \aj , 124, 2625
\reference{met} Saviane, I., Held, E.V., Bertelli, 2000, A\&A, 355, 56
\reference{HRCARINA} Smecker-Hane, T.A., Mandushev, G.I., Hesser, J.E., 
Stetson, P.B., Da Costa, G.S., Hatzidimitriou, D., 1999, in 
ASP Conf. Ser., vol 192, p. 159
\reference{ext} Schlegel, D.J., Finkbeiner, D.P.,  and Davis, M., 1998,
\apj, 500, 525
\reference{met} Tolstoy, E., et al., 2002, Astroph. and Space Science, 281, 217
\reference{clust} Udalski, A., 1998, Acta Astron., 48, 383
\reference{rcogle} Udalski, A., 2000a, \apj, 531, L25
\reference{diff} Udalski, A., 2000b, Acta Astron., 50, 279
\reference{SMC} Udalski, A., Szyma\'nski, M., Kubiak, M., Pietrzy\'nski, G.,
Wo\'zniak, P., and \.Zebru\'n, K. 1998,  Acta Astron., 48, 147
\reference{reden-lmc} Udalski, A., Soszy{\'n}ski, I.,  Szyma\'nski, M., Kubiak, M.,
 Pietrzy\'nski, G, Wo\'zniak, P., and \.Zebru\'n, K. 1999a,  Acta Astron., 49, 223
\reference{reden-smc} Udalski, A., Soszy{\'n}ski, I.,  Szyma\'nski, M., Kubiak, M.,
 Pietrzy\'nski, G, Wo\'zniak, P., and \.Zebru\'n, K. 1999b,  Acta Astron., 437, 520
\reference{LMC} Udalski, A., Szyma\'nski, M., Kubiak, M., Pietrzy\'nski, G.,
Wo\'zniak, P., and \.Zebru\'n, K. 2000,  Acta Astron., 50, 307
\reference{mar} van der Marel, R.P., Alves, D.R., Hardy, E., and Suntzeff, N.B., 2002,
astro-ph/0205161   
\end{references}
\end{document}